# Modeling, Simulation and Implementation of a Bird-Inspired Morphing Wing Aircraft

Kun Xiao[1], Yuxin Chen[2], Wuyao Jiang[2], Chenyao Wang[2], Longfei Zhao[1]
[1] School of Automation Science and Electrical Engineering, Beihang University
Beijing, China
[2] School of Aeronautic Science and Engineering, Beihang University
Beijing, China
e-mail: zhaolf@buaa.edu.cn

*Abstract*—We present a design of a bird-inspired morphing wing aircraft, including bionic research, modeling, simulation and flight experiments. Inspired by birds and activated by a planar linkage, our proposed aircraft has three key states: gliding, descending and high-maneuverability. We build the aerodynamic model of the aircraft and analyze its mechanisms to find out a group of optimized parameters. Furthermore, we validate our design by Computational Fluid Dynamics (CFD) simulation based on Lattice-Boltzmann technology and determine three phases of the planar linkage for the three states. Lastly, we manufacture a prototype and conduct flight experiments to test the performance of the aircraft.

## I. Introduction

With the rapid development of aeronautics, fixed-wing aircrafts have surpassed birds in many aspects. However, considering the multi-mission adaptive flying, fixed-wing aircrafts still fall behind birds, because fixed wings are only able to bring the aircraft good performance during a specific flight mission. For example, the aircraft with large aspect ratio such as Northrop Grumman RQ-4 Global Hawk perform well at a subsonic speed, and the aircraft with small aspect ratio such as General Dynamics F-16 Fighting Falcon, do well in a supersonic flight. Morphing wing technique is used to improve performance in multi-mission flight.

A morphing wing aircraft is an aircraft which changes the shape of wing during flight in different respects, such as area of wing, aspect ratio, angle of swept back and so on, so that the aircraft can keep the best performance in different flight missions [1]. At present, some research institutions have made some progress in this direction. NASA Langley Research Center contributes to research into smart materials and new actuators for morphing wing aircraft [2,3]. The Pentagon Defense Pre-Research Program and the Air Force Research Laboratory have implemented the Morphing Aircraft Structure (MAS) program [4]. Furthermore, MIT and NASA are cooperating to design a brand-new efficient morphing wing [5].

Birds are good examples of morphing wing aircrafts in nature. Birds have the ability to change the aspect ratio, airfoil and dihedral angle. When flying at low speeds, birds stretch their wings to increase the lift-to-drag ratio. And when flying at high speeds, birds fold their wings to reduce the lift-to-drag ratio. Birds can also increase the dihedral angle during the descent to increase the lateral stability [6,7]. Figure 1 shows the wing morphology of an owl when it is gliding and descending. It can be seen that when soaring the owl stretches its wing to increase the aspect ratio and when descending it increases the dihedral angle.

Inspired by morphing wings of birds, we designed a bird-like morphing wing aircraft, which is shown as Figure 2. It is worth noting that, our design is quite different from ornithopters, which is another branch of aeronautics. A morphing wing aircraft is still a fixed-wing aircraft. The largest difference between it and an ornithopter is that the thrust does not come from flapping wings but from other dynamic systems such as a motor-propeller system or a jet engine. Flapping wings lead to strong unsteady aerodynamics [8] but morphing wings do not. A morphing wing aircraft in a static configuration has approximately steady aerodynamics, just like a conventional fixed-wing aircraft, but in the process of morphing, its aerodynamics is unsteady, which results in difficulty in control [9]. In this paper, only the steady states of morphing wing are considered, and the solution to the control problem caused by the process of wing morphing is our future work.

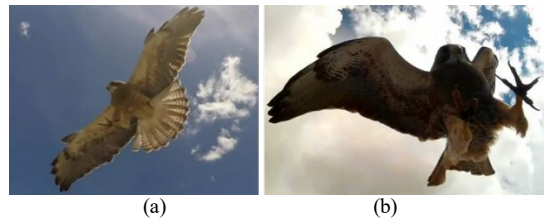

Figure 1. Wing morphology og owl: (a) gliding (b) descemding

Our proposed bird-inspired morphing wing aircraft has a pair of outer wings and inner wings. Its morphing motion is activated by a large power servo and two small power servos. The large power servo activates a mechanical linkage to adjust the aspect ratio and the dihedral angle, and two small power servos adjust the twist distribution of the outer wings. There are three main flight states, which are gliding, descending and high-maneuverability. The features of the three flight states are listed as Table 1 and the 3D models are shown as Figure 2

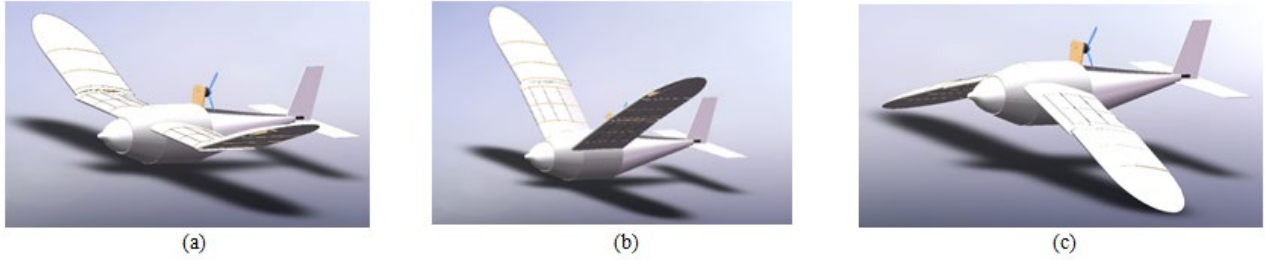

Figure 2. 3D models of the aircraft in three flight states: (a) gliding (b) descemding (c) high-maneuverability

TABLE I. FEATURES OF THREE FLIGHT STATES

|  | *Lift-drag ratio* | *Lateral stability* |
| --- | --- | --- |
| Gliding | High | Middle |
| Descending | Low | High |
| High-maneuverability | Middle | Instability |

Our work includes bionic research, modeling, simulation and flight experiments. Firstly, we choose the planar linkage mechanism as the main mechanism by analyzing the motion of large birds. Secondly, we find out a group of optimized parameters based on the aerodynamic model and mechanisms. Thirdly, we validate our design by the way of Computational Fluid Dynamics (CFD) simulation based on Lattice-Boltzmann technology and determine three phases of the linkage mechanism for gliding state, descending state and high-maneuverability state. Lastly, we manufacture a prototype and conduct flight experiments to validate the performance of our bionic morphing wing aircraft.

## II. MOTION ANALYSIS OF LARGE BIRDS

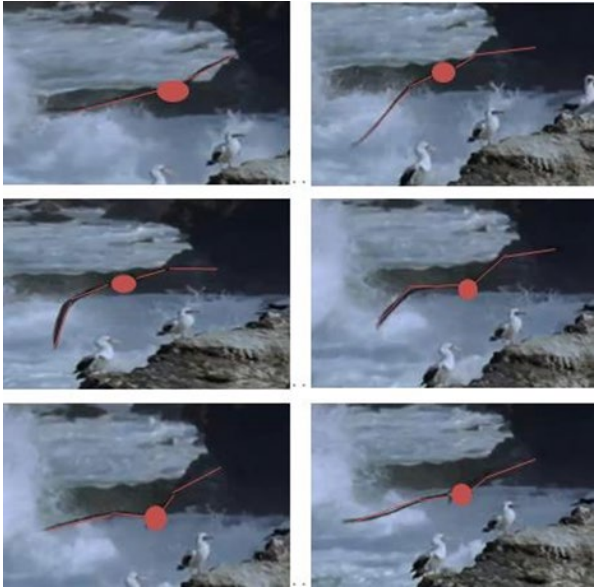

Figure 3. Flight analysis of an albatross

There are numerous kinds of birds, from which we choose large bird species as research objects. Their features are large wing span, frequent gliding aided with low-frequency flapping. Therefore, the wing morphing of large birds is highly valuable for the design of morphing wing aircrafts. By analyzing the video of the flight of an albatross in Figure 3, we abstract a motion model, which is showed as Figure 4.

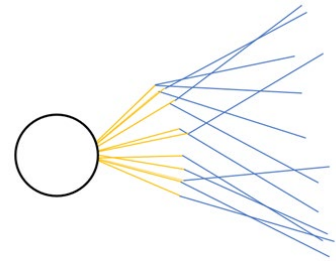

Figure 4. Wing motion model (for one side) of an albatross

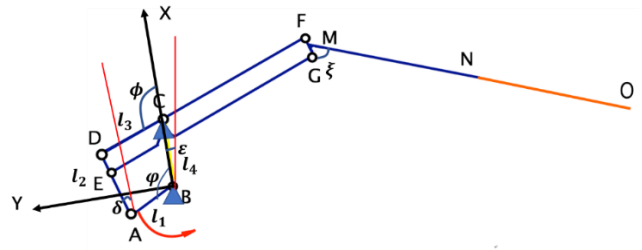

Figure 5. Planar linkage mechanism (for one side)

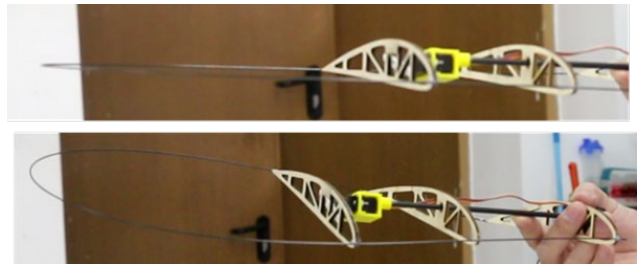

Figure 6. Twist of an outer wing

To realize the similar motion in Figure 3, we designed a planar linkage mechanism shown as Figure 5. NO is a flexible linkage made by carbon fiber and polypropylene, polyethylene, shown in Figure 6 and Figure 12. The flexible design is used to make the twist distribution of outer wings adjustable, which is another bird-inspired design. Wing twist can adjust lift distribution along wing span and prevent wing stall. Additionally, for our proposed aircraft, the two servo-activated twistable outer wings are regarded as flaperons.

### III. AERODYNAMIC MODELING AND DESIGN

In order to determine the lengths of the linkages, an aerodynamic model of the morphing wing is built.

#### A. Wing Shape and Airfoil

According to the basic research of aerodynamics, elliptical wings have the best stall characteristics [10], while rectangular wings are the simplest in terms of processing technology. The proposed aircraft controls the angle of attack through the torsion of the outer wing, so the shape of outer wing is designed as half an ellipse, which is shown in Figure 2, to realize synchronous stall.

The proposed aircraft requires an airfoil with a large stall angle, a high lift-drag ratio and a small pitch torque coefficient. In Figure 7, five commonly used airfoils are compared by using *Profili*[1], and airfoil USA35-B is selected for its good overall performance.

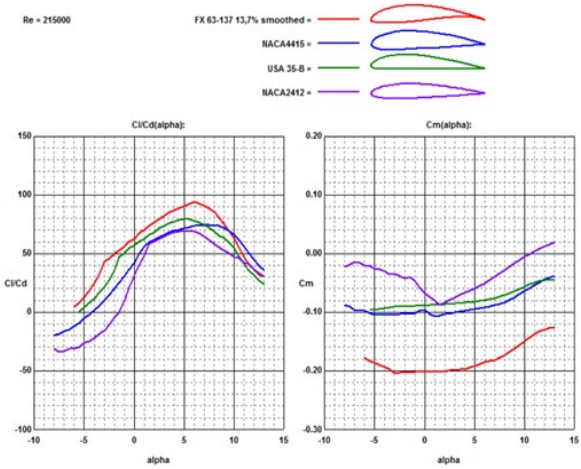

Figure 7. Aerodynamics characteristics of five airfoils

#### B. Lift-drag Ratio

The aerodynamic characteristics and flight dynamics characteristics of the design are analyzed in air-path reference frame ($Ox_a y_a z_a$) and body reference frame ($Ox_b y_b z_b$) (see Figure 8)[11].

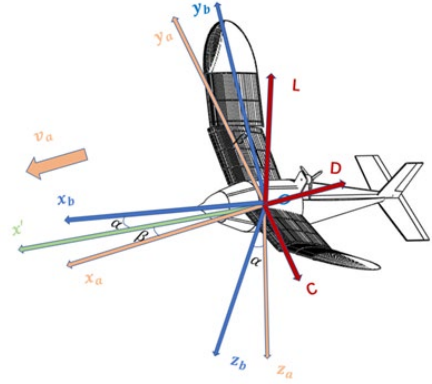

Figure 8. Air-path reference frame and body reference frame

The lift L and drag D of the wings of the aircraft are described as follows:

$$L = \frac{1}{2}\rho\|v_a\|^2 S C_L \quad (1)$$

$$D = \frac{1}{2}\rho\|v_a\|^2 S C_D \quad (2)$$

where S is wing area, $v_a$ is air speed, $\rho$ is density, $C_L$ is lift coefficient and $C_D$ is drag coefficient [12].

According to Anderson et al. [13], the lift coefficient can be calculated by Eq. (3) within the stall angle

$$C_L = C_L^\alpha (\alpha - \alpha_0) \quad (\alpha \leq \alpha_s) \quad (3)$$

where $\alpha$ is the angle of attack, and $C_L^\alpha$, $\alpha_0$ and $\alpha_s$ are determined by the airfoil.

According to Cone et al. [14], the total drag coefficient for a wing can be written as

$$C_D = C_{D0} + C_{Di} = C_{D0} + A C_L^2 \quad (4)$$

where $C_{D0}$ is defined as the drag coefficient of zero-lift, $C_{Di}$ is the drag coefficient due to lift, and A is the factor of the drag coefficient due to lift. $C_{D0}$ and A are functions of Mach number Ma and Reynolds number Re, but within the low speed ($Ma \leq 0.3$) $C_{D0}$ and A can be considered as constants, usually as 0.02 and 0.2 respectively.

From Eq. (1) and Eq. (2), the lift-to-drag ratio K can be defined as

$$K = \frac{L}{D} \quad (5)$$

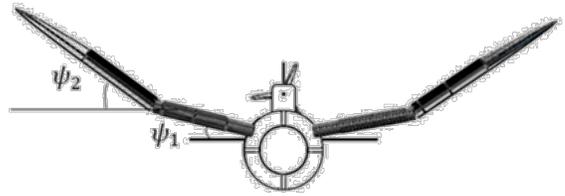

Figure 9. Angles of the inner and outer wings

In order to calculate the lift of the entire wing, the angles of the inner and outer wings are defined as $\psi_1$ and $\psi_2$, which are positive when the chord of the wings is above the Oxy plane (see Figure 8).

---
[1] http://www.profili2.com/

According to Eq. (1), when the roll motion is not taken into account, total lift and drag of the wing can be calculated by Eq. (6) and Eq. (7)

$$L = \frac{1}{2}\rho U^2 C_L (S_1 \cos \psi_1 + S_2 \cos \psi_2) \times 2 \quad (6)$$

$$D = \frac{1}{2}\rho U^2 C_D (S_1 + S_2) \times 2 \quad (7)$$

where $S_1$ is the area of the inner wing, and $S_2$ is the area of outer wing.

From Eq. (5), Eq. (6) and Eq. (7), the lift-to-drag ratio K is:

$$K = \frac{C_L}{C_D} \frac{S_1 \cos \psi_1 + S_2 \cos \psi_2}{S_1 + S_2} \quad (8)$$

From Eq. (8), it is clear that the lift-to-drag ratio depends on $\frac{S_1 \cos \psi_1 + S_2 \cos \psi_2}{S_1 + S_2}$. If $S_1$ and $S_2$ are fixed, the closer the $\psi_1$ and $\psi_2$ are to 0°, the higher the K is; if $\psi_1$ and $\psi_2$ are fixed and $\frac{\cos \psi_2}{\cos \psi_1} < 1$, the higher $\frac{S_1}{S_2}$ is, the higher K is; if $\psi_1$ and $\psi_2$ are fixed and $\frac{\cos \psi_2}{\cos \psi_1} > 1$, the higher $\frac{S_1}{S_2}$ is, the lower K is.

### C. Lateral Stability

The roll caused by a disturbance from the symmetric equilibrium condition also generates forces and moments. If the roll angle is negative, the resultant lift and the gravity will cause a leftward sideslip. If a roll moment is generated to return to the horizontal state, the aircraft will possess the lateral stability.

The critical derivative of the lateral stability is defined as $C_{l\beta} = \frac{\partial C_l}{\partial \beta}$ ($C_l$ is the roll moment coefficient and $\beta$ is the angle of sideslip) and when it is negative the aircraft is stable laterally. The total $C_{l\beta}$ is

$$C_{l\beta} = C_{l\beta,w} + C_{l\beta,f} + C_{l\beta,vt} \quad (9)$$
$$C_{l\beta,w} = C_{l\beta,0°} + C_{l\beta,\psi} \quad (10)$$

where $w, f, vt$ represents wing, fuselage and vertical tail respectively. $0°$ means that the wings are non-swept and $\psi$ means the dihedral angel of the wings [16]. The contribution to the derivative from the dihedral angle is

$$C_{l\beta,\psi} = -\frac{1}{2} C_{L\alpha} \overline{y_{s,c}} \psi \quad (11)$$

Thus, the total contribution of the wings is

$$C_{l\beta,\psi} = -\frac{1}{2} C_{L\alpha} \overline{y_{s,c1}} \psi_1 - \frac{1}{2} C_{L\alpha} \overline{y_{s,c2}} \psi_2 \quad (12)$$

Since the tip is elliptical, the contribution from the tip effect $C_{l\beta,0°}$ can be neglected. Generally, $C_{l\beta,vt}$ is negative for the conventional tails [17]. The contribution from the fuselage can be neglected when the mass center is on the longitudinal symmetrical plane, which means $C_{l\beta,f} \approx 0$[18].

Above all, if $C_{l\beta,\psi} < 0$, the aircraft will possess lateral stability. Because $C_{l\beta}$ is dependent on $C_{l\beta,\psi}$, the smaller $C_{l\beta,\psi}$ is, the more stable the aircraft will be. On the contrary, a larger $C_{l\beta,\psi}$ will yield a better lateral maneuverability.

### D. Three Flight States

Combining the requirements of aerodynamic characteristics and stability discussed above with the bionic study of birds, we select three main flight states:

a. Gliding
This state requires a high lift-to-drag ratio K to generate lift and a middle lateral stability to maintain the flight attitude.
b. Descending
This state requires the highest stability and the lowest lift-to-drag ratio.
c. High-maneuverability
This state demands a high maneuverability, or a certain degree of instability, and can sacrifice part of the aerodynamic characteristics for a middle K.

The characteristic parameters $\psi_1$ and $\psi_2$ of the three states are chosen as follows

TABLE II. CHARACTERISTIC PARAMETERS OF THREE STATES

| | $\psi_1$[deg] | $\psi_2$[deg] |
|---|---|---|
| Gliding | -2~2 | 20~30 |
| Descending | 35~45 | Same as $\psi_1$ |
| High-maneuverability | -30~-20 | Same as $\psi_1$ |

### E. Mechanism Design

According to the reference range of $\psi_1$ and $\psi_2$ in Table 2, we designed a planar linkage mechanism (see Figure 5).

The mechanism itself has the following constraints [19, 20]

$$\cos \varphi = l_3 \cos \phi - \frac{l_3}{l_4} \cos(\phi - \varphi) + \frac{l_4^2 + l_3^2 + l_1^2 - l_2^2}{2l_4} \quad (13)$$

Variables $\varphi, \phi$ and parameters $l_1, l_2, l_3, l_4, \varepsilon, \xi$ are shown in Figure 5 and subscripts [1], [2], [3] in Eq. 14 stand for states of gliding, descending and high-maneuverability respectively.

$$\begin{cases} \cos \varphi_{[1]} = l_3 \cos \phi_{[1]} - \frac{l_3}{l_4} \cos(\phi_{[1]} - \varphi_{[1]}) + \frac{l_4^2 + l_3^2 + l_1^2 - l_2^2}{2l_4} \\ \cos \varphi_{[2]} = l_3 \cos \phi_{[2]} - \frac{l_3}{l_4} \cos(\phi_{[2]} - \varphi_{[2]}) + \frac{l_4^2 + l_3^2 + l_1^2 - l_2^2}{2l_4} \\ \cos \varphi_{[3]} = l_3 \cos \phi_{[3]} - \frac{l_3}{l_4} \cos(\phi_{[3]} - \varphi_{[3]}) + \frac{l_4^2 + l_3^2 + l_1^2 - l_2^2}{2l_4} \end{cases} \quad (14)$$

The relationship between $\psi_1, \psi_2$ and $\varphi, \phi, \varepsilon, \xi$ is shown in Eq. (15)

$$\begin{cases} \phi - \psi_1 = 90° - \varepsilon \\ \xi - \psi_2 = 90° - \varepsilon \end{cases} \quad (15)$$

Constraints in Eq. 14, Eq. 15 and Table 2 are insufficient for the calculation of the specific values of each variable and the parameters are still unknown. Considering structural strength and transmission angle, the values of parameters are determined by the cut-and-try method, as is shown in Table 3.

TABLE III. VALUES OF PARAMETERS

| $l_1$[mm] | 26.2 | EG[mm] | 265.6 |
|---|---|---|---|
| $l_4$[mm] | 52.2 | CF[mm] | 220.6 |
| $l_3$[mm] | 46.9 | MN (outer wing)[mm] | 178.2 |
| $l_2$[mm] | 45.6 | $\xi$[deg] | 60 |
| DE[mm] | 14.2 | $\varepsilon$[deg] | 21.24 |

## IV. SIMULATION AND FLIGHT EXPERIMENT

### A. CFD Simulation

To determine the three phases of the linkage mechanism for gliding state, descending state and high-maneuverability state, a Computational Fluid Dynamics (CFD) simulation is conducted. In this research, *XFlow* [21], which offers particle-based Lattice-Boltzmann CFD applications, is used as the simulated wind tunnel experiment platform. The key simulation settings are listed in Table 4. Figure 10 shows the pressure field.

TABLE IV. KEY SIMULATION SETTINGS

| Wind tunnel size[$m^3$] | $x*y*z=$ $10*4*4$ | Basic resolution[m] | 0.01 |
|---|---|---|---|
| Airspeed[m/s] | 10 | Shape refinement resolution[m] | 0.0015625 |
| Angle of attack [deg] | 5 | Wake resolution[m] | 0.0015625 |
| Angle of sideslip[deg] | 5 | | |

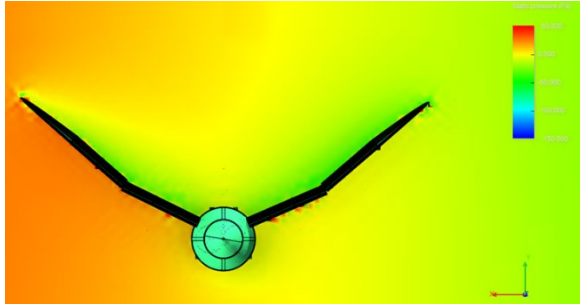

Figure 10. Pressure field

After data post-processing [22], a lift-drag ratio curve and a roll moment curve are obtained, as is shown in Figure

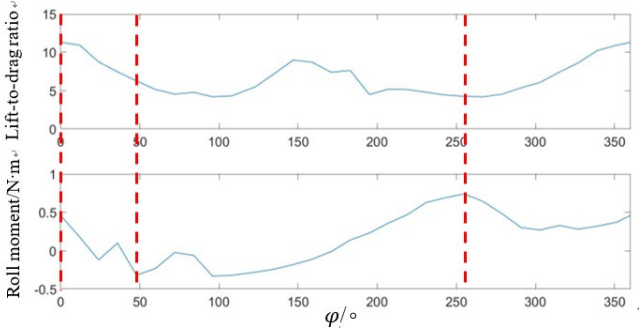

Figure 11. A Lift-drag ratio curve(above) and roll moment curve(below)

According to Table 1, gliding state needs high lift-drag ratio and middle lateralstability, descending state needs low lift-to-drag ratio and high lateral stability, and high-maneuverability state needs middle lift-drag ratio and lateral instability. By combining the two curves, three phases (described by φ) of the planar linkage are determined. Thus, dihedral angles of the inner wing and the outer wing, lift-to-drag ratios and lateral stabilities of three flight states are determined, which are shown as Table 5.

TABLE V. KEY VALUES OF THREE FLIGHT STATES

| Flight state | Phase [deg] | dihedral angles of the inner wing[deg] | dihedral angles of the outer wing [deg] | Lift-drag ratio | Roll moment [N·M] |
|---|---|---|---|---|---|
| Gliding | 0 | -1 | 27.1 | 11.3 | 0.46 |
| Descending | 48 | 37.8 | 38.1 | 6.3 | -0.32 |
| High-maneuverability | 256 | -21.4 | -23.2 | 4.26 | 0.74 |

### B. Experiment

For morphing wing aircraft, great importance is attached to the manufacturing process, especially the choice of materials. To obtain high structural strength and stiffness, low mass, as well as flexibility of skin, carbon fiber, wood and plastic are used.

According to the design discussed above, a prototype is built, which is shown as Figure 12. Table 6 lists the parameters of the prototype.

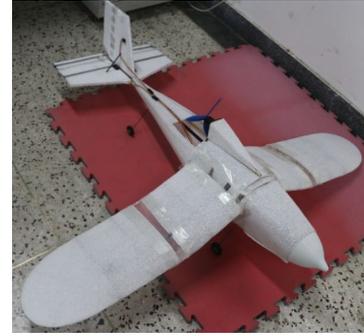

Figure 12. Prototype

TABLE VI. PARAMETERS OF THE PROTOTYPE

| Weight[g] | 1224 | structure | Carbon fiber, wood, polylactide |
|---|---|---|---|
| Wingspan[m] | 1.35 | Skin | Polypropylene, polyethylene |
| Torso length[m] | 1.00 | Battery | Lithium polymer accumulator, 3cells, 2200 mAh |
| Propeller | 6040 | Motor | Brushless 2216, 1800KV |

In order to test the performance and reliability of the design, ground experiments and flight experiments are conducted respectively. The prototype performs well in the gliding state and the descending state. Because the current research does not include control system, which involves unsteady aerodynamics, the high-maneuverability flight experiment was not conducted.

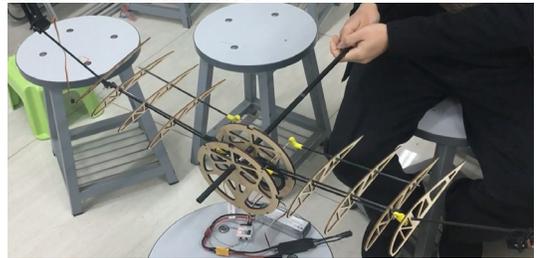

(a)

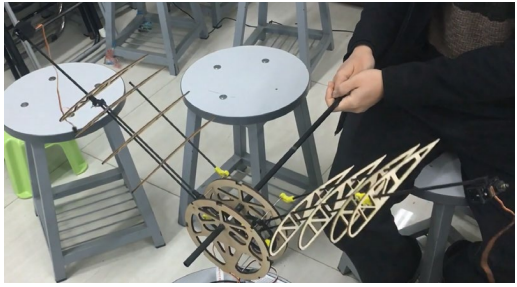
(b)

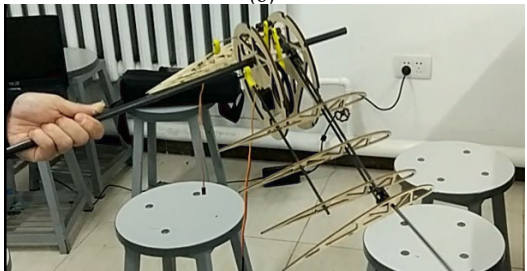
(c)

Figure 13. Ground experiments: (a) Gliding state (b) Descending state (c) High-maneuverability state

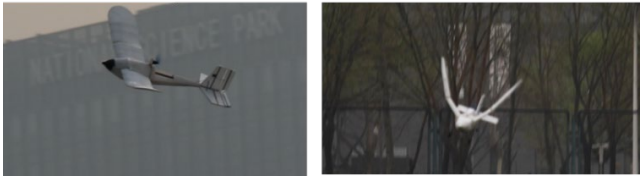
(a)                      (b)

Figure 14. Flight experiments: (a) Gliding state (b) Descending state

## V. CONCLUSION

A design of a bird-inspired morphing wing aircraft is presented. Based on the motion analysis of large birds, we choose the planar linkage mechanism as the main mechanism to realize three flight states, gliding, descending and high-maneuverability. To find out a group of optimized parameters for it, we build an aerodynamic model and analyze the mechanisms. Furthermore, we validate our design in XFlow, a CFD simulation platform based on Lattice-Boltzmann technology and determine three phases of the linkage mechanism for the three states. Lastly, we manufacture a prototype and conduct flight experiments to test the performance of our designed bionic morphing wing aircraft.

Our future work focuses on the process of wing morphing, which has unsteady aerodynamics. And based on it, a controller for the morphing wing aircraft will be designed, which will make the high-maneuverability flight discussed above achievable.